\documentclass[12pt]{article}
\usepackage{amssymb,amsmath,amsfonts,amsthm,amstext,amscd,array}
\usepackage{mathrsfs}
\usepackage{hyperref}
\usepackage{pdfsync}
\usepackage{bbm}
\usepackage{bm}
\usepackage[arrow,matrix,curve]{xy}
\usepackage{bbding}
\usepackage{wasysym}

\usepackage{epsf}
\usepackage{epsfig}
\usepackage{wrapfig}
\input{epsf}

\usepackage{tempora}
\usepackage{color,xcolor}
\usepackage{manfnt}

\begin{document}

\title{\Large\bf  Canonical representation of the Snyder-de Sitter algebra with correct flat and commutative limits}

\author{
V.G. Kupriyanov$^{1}$ and E.L.F. de Lima$^{1}$
\\[2mm]
{\small $^{1}$Centro de Matemática, Computação e Cognição,
Universidade Federal do ABC, Santo André, Brazil}
\\
{\small Emails: \texttt{vladislav.kupriyanov@gmail.com},
\texttt{eduardo.lourenco@ufabc.edu.br}}
}
\maketitle
%\affiliation{CMCC - Universidade Federal do ABC - Brazil}
\begin{abstract}
The Snyder--de Sitter algebra provides a Lorentz-covariant deformation of
phase-space geometry characterized by a curvature parameter $\alpha$ and a
noncommutativity parameter $\beta$. We construct an explicit canonical
(Darboux) representation of this algebra that is regular in both
parameters. Starting from the symplectic structure associated with the
Snyder--de Sitter Poisson brackets, we derive the canonical
transformation between the physical phase-space variables and Darboux
coordinates and obtain its inverse in closed form to all orders in
$\alpha$ and $\beta$. The resulting representation has well-defined flat ($\alpha\to0$) and
commutative ($\beta\to0$) limits, reducing respectively to the Snyder and
de Sitter phase-space algebras, while the simultaneous limit
$(\alpha,\beta)\to(0,0)$ yields the standard canonical phase-space
coordinates.
We then apply this representation to the construction of Poisson gauge
transformations on Snyder--de Sitter phase space. In particular, we derive
the corresponding gauge transformation matrix and Poisson field strength,
both of which are regular in $\alpha$ and $\beta$. Our construction
provides a convenient framework for investigating gauge theories and
other physical systems formulated on Snyder--de Sitter phase space.
\end{abstract}

\section{Introduction}

Noncommutative geometry has long been regarded as a natural mathematical
framework for describing possible Planck-scale modifications of spacetime.
Several proposals for noncommutative algebras of spacetime coordinates have
been considered in the literature. In this work, we are particularly
interested in models that preserve Lorentz invariance. The Snyder-de Sitter
(SdS) algebra of Poisson brackets, which originally appeared in the context
of Triply Special Relativity \cite{Kowalski-Glikman:2004fso}, is given by
\begin{eqnarray}\label{SdS1}
\{x_\mu,x_\nu\}=\beta^2\,M_{\mu\nu}\,,\qquad
\{p_\mu,p_\nu\}=\alpha^2\,M_{\mu\nu}\,,
\end{eqnarray}
with
\begin{eqnarray}
&&\{M_{\mu\nu},M_{\rho\sigma}\}
=\eta_{\mu\rho}\,M_{\nu\sigma}
-\eta_{\mu\sigma}\,M_{\nu\rho}
-\eta_{\nu\rho}\,M_{\mu\sigma}
+\eta_{\nu\sigma}\,M_{\mu\rho}\,,\label{M1}\\
&&\{M_{\mu\nu},x_\rho\}
=\eta_{\mu\rho}\,x_\nu-\eta_{\nu\rho}\,x_\mu\,,
\qquad
\{M_{\mu\nu},p_\rho\}
=\eta_{\mu\rho}\,p_\nu-\eta_{\nu\rho}\,p_\mu\,.\label{M2}
\end{eqnarray}
Here, $\alpha$ and $\beta$ are two real parameters and
$\eta_{\mu\nu}$ is the Minkowski metric. In addition,
\begin{equation}\label{gSg}
\{x_\mu,p_\nu\}
=\eta_{\mu\nu}
+\alpha^2x_\mu\,x_\nu
+\beta^2p_\mu\,p_\nu
+2\alpha\beta\,x_\nu\,p_\mu\,,
\end{equation}
and the Lorentz generators are given by
\begin{equation}\label{Lorentzgen}
M_{\mu\nu}=x_\mu\,p_\nu-x_\nu\,p_\mu\,.
\end{equation}

It has been conjectured that this algebra may be relevant for the low-energy
behavior of quantum gravity in the presence of a nonzero cosmological
constant. In the limit $\beta\to0$, it reduces to the phase-space algebra of
a particle in de Sitter spacetime. We therefore interpret $\alpha$ as a
curvature parameter. On the other hand, setting $\alpha\to0$ yields the
Snyder algebra \cite{Snyder:1946qz}, describing a noncommutative spacetime
that is manifestly Lorentz covariant. Accordingly, $\beta$ is interpreted as
the noncommutativity parameter.

Remarkably, this algebra enjoys Born reciprocity \cite{Born:1949yva}:
the simultaneous transformations $x\leftrightarrow p$ and
$\alpha\leftrightarrow\beta$ leave the algebra invariant. Moreover, the
Snyder-de Sitter algebra is the simplest known algebra with this property,
as it is quadratic in both $x$ and $p$. Specific realizations of the Yang
algebra \cite{Yang:1947ud}, which also enjoys Born reciprocity, were
constructed in \cite{Guo:2008qp,Bilac:2024wex,Bilac:2024xxq,Bilac:2026vef}
and are considerably more complicated than (\ref{gSg}).

Our aim in this work is to construct a canonical representation of the
algebra (\ref{SdS1})--(\ref{Lorentzgen}) that is regular in both parameters
$\alpha$ and $\beta$. More precisely, let
$Y^{\mathcal M}=(y^\mu,\xi_\mu)$, $\mathcal M=1,\dots,2d$, be a set of
canonical (Darboux) phase-space coordinates satisfying
\begin{equation}\label{Dc2na}
\{y^\mu,y^\nu\}=0\,,\qquad
\{y^\mu,\xi_\nu\}=\delta^\mu_\nu\,,\qquad
\{\xi_\mu,\xi_\nu\}=0\,.
\end{equation}
We seek functions $x_\mu=x_\mu(y,\xi)$ and
$p_\mu=p_\mu(y,\xi)$ such that
\begin{equation}\label{regular}
\lim_{(\alpha,\beta)\to(0,0)}x_\mu(y,\xi)=y_\mu
\qquad\mbox{and}\qquad
\lim_{(\alpha,\beta)\to(0,0)}p_\mu(y,\xi)=\xi_\mu\,.
\end{equation}

Canonical representations of the Snyder-de Sitter algebra have previously
been constructed, e.g., in \cite{Mignemi:2011wh,Mignemi:2015una}.
However, none of the known expressions satisfies the regularity condition
(\ref{regular}). Perturbative expressions were obtained in
\cite{Meljanac:2022qhp}, but an explicit all-order expression was not
provided. The paper is organized as follows. In Sec.~2, we construct the
action principle $S[X]$, with $X^{\mathcal M}=(x^\mu,p_\mu)$, that yields
the Poisson brackets defined in (\ref{SdS1})--(\ref{Lorentzgen}). In Sec.~3, we consider the action $S[Y]$ reproducing the canonical
brackets (\ref{Dc2na}) and construct the change of variables $Y=Y(X)$
relating the two action principles. In Sec.~4, we derive the inverse
transformation $X=X(Y)$, which constitutes the main technical result of
this work. Finally, in Sec.~5, we discuss the application of the
resulting canonical representation to the construction of the Poisson
gauge algebra on Snyder--de Sitter space.

\section{Action principle}

The Snyder--de Sitter Poisson algebra can be written in the condensed form
\begin{eqnarray}\label{SdS}
&&\{X^{\cal M},X^{\cal N}\}=\Theta^{{\cal M}{\cal N}}(X)\,,\\
&&\Theta^{{\cal M}{\cal N}}=
\left({\begin{array}{cc}
\beta^2\left(x^\mu\,p^\nu-x^\nu\,p^\mu\right)
&
\delta^\mu_\nu+\alpha^2x^\mu\,x_\nu+\beta^2p^\mu\,p_\nu
+2\alpha\beta\,p^\mu \,x_\nu\\
-\delta^\nu_\mu-\alpha^2x^\nu\,x_\mu-\beta^2p^\nu\,p_\mu
-2\alpha\beta\,p^\nu \,x_\mu
&
\alpha^2\left(x_\mu\,p_\nu-x_\nu\,p_\mu\right)
\end{array}}\right).
\notag
\end{eqnarray}

The inverse of the Poisson tensor defines the corresponding symplectic
two-form
\begin{equation}\label{sf}
\Omega=\frac{1}{2}\,\Theta^{-1}_{{\cal M}{\cal N}}\,
dX^{\cal M}\wedge dX^{\cal N},
\end{equation}
where
\begin{eqnarray}\label{theta-inv}
&&\Theta^{-1}_{{\cal M}{\cal N}}=
\frac{1}{1+\lambda}\times\\
&&\left({\begin{array}{cc}
\alpha^2 \left(x_\mu\,p_\nu-x_\nu\,p_\mu\right)
&
-\delta^\nu_\mu\left(1+\lambda\right)
+\alpha^2x^\nu\,x_\mu+\beta^2p^\nu\,p_\mu
+2\alpha\beta\,p^\nu\,x_\mu
\\
\delta^\mu_\nu\left(1+\lambda\right)
-\alpha^2x^\nu\,x_\mu-\beta^2p^\nu\,p_\mu
-2\alpha\beta\,p^\mu\,x_\nu
&
\beta^2 \left(x^\mu\,p^\nu-x^\nu\,p^\mu\right)
\end{array}}\right),
\notag
\end{eqnarray}
and
\begin{equation}
\lambda=\alpha^2\,x^2+\beta^2\,p^2
+2\,\alpha\,\beta\,x\cdot p\,.
\end{equation}

To construct the symplectic potential
$J=J_{\cal M}(X)\,dX^{\cal M}$, with
$J_{\cal M}(X)=(j_\mu(x,p),j^\mu_\ast(p,x))$, whose exterior derivative
reproduces the two-form (\ref{sf}), we use the ansatz
\begin{eqnarray}\label{a1}
j_\mu&=&f\left(u,v,z\right)x_\mu+g\left(u,v,z\right)p_\mu\,,
\nonumber\\
j^\mu_\ast&=&h\left(u,v,z\right)x^\mu
+k\left(u,v,z\right)p^\mu\,,
\end{eqnarray}
where
$u=x^2/2$, $v=p^2/2$, and $z=x\cdot p$.
With this ansatz, both $j_\mu$ and $j^\mu_\ast$ transform as Lorentz
vectors, ensuring the Lorentz invariance of the corresponding action.
Imposing the condition $dJ=\Omega$ and using the explicit form of
$\Theta^{-1}$ in (\ref{theta-inv}), we obtain
\begin{equation}\label{g-h}
g-h=1\,,
\end{equation}
together with
\begin{eqnarray}
\label{eq:fghk-closed}
f&=&-2(\alpha^2z+2\alpha\beta v)F(\lambda), \nonumber\\
g&=&\frac12+2(\alpha^2u-\beta^2v)F(\lambda), \nonumber\\
k&=&2(\beta^2z+2\alpha\beta u)F(\lambda),
\end{eqnarray}
where
\begin{equation}
\label{eq:F-Lambda-definition}
F(\lambda)=\frac{\lambda-\ln(1+\lambda)}{2\lambda^2}
\end{equation}
is regular at $\lambda=0$, with $F(0)=1/4$.

The corresponding action functional is
\begin{equation}\label{ac1}
S=\int d\tau\left[J_{\cal M}(X)\,\dot X^{\cal M}-H(X)\right],
\end{equation}
where $H(X)$ is an arbitrary Hamiltonian. We also note that the Poisson
structure (\ref{SdS}) can be obtained from a higher-dimensional model with
canonical Poisson structure by imposing suitable constraints
\cite{Carrisi:2010jv}.

\section{Change of variables}

Consider the set of canonical (Darboux) variables
$Y^{\cal M}=(y^\mu,\xi_\mu)$ satisfying the Poisson brackets
(\ref{Dc2na}). These brackets follow from the action
\begin{equation}\label{ac2}
S_0=\int d\tau\left[
\frac{1}{2}\Big(\xi_\mu\,\dot{y}^\mu-y^\mu\,\dot{\xi}_\mu\Big)
-H_0(Y)\right],
\end{equation}
where we have chosen the symmetric symplectic potential
$J^0_{\cal M}(Y)=\frac12(\xi_\mu,-y^\mu)$.

Our aim is to express the original phase-space coordinates
$X^{\cal M}=(x^\mu,p_\mu)$, which satisfy the Poisson brackets
(\ref{SdS}), in terms of the canonical variables $Y^{\cal M}$.
Equivalently, we first construct the inverse transformation
$Y=Y(X)$. Performing this change of variables in (\ref{ac2}) and
comparing the resulting action with (\ref{ac1}), we obtain
\begin{equation}\label{eq1}
\xi_\nu\,\partial_\mu y^\nu-y^\nu\,\partial_\mu \xi^\nu
=2\,j_\mu(x,p)\,,
\qquad
\xi_\nu\,\partial^\mu_p y^\nu-y^\nu\,\partial^\mu_p\xi_\nu
=2\,j^\mu_\ast(x,p)\,,
\end{equation}
where
$\partial_\mu=\partial/\partial x^\mu$ and
$\partial^\mu_p=\partial/\partial p_\mu$, while the functions
$j_\mu(x,p)$ and $j^\mu_\ast(p,x)$ are determined by (\ref{a1}).

Again, both $y^\mu$ and $\xi_\mu$ should transform as Lorentz vectors.
We therefore consider the ansatz
\begin{equation}\label{can}
y^\mu=a(u,v,z)\,x^\mu+b(u,v,z)\,p^\mu\,,
\qquad
\xi_\mu=c(u,v,z)\,x_\mu+d(u,v,z)\,p_\mu\,.
\end{equation}

In terms of the canonical variables, the Lorentz generators are
\[
M^{\mu\nu}=y^\mu\xi^\nu-y^\nu\xi^\mu.
\]
On the other hand, in terms of the original phase-space variables they
are given by (\ref{Lorentzgen}),
\[
M^{\mu\nu}=x^\mu p^\nu-x^\nu p^\mu.
\]
Consistency of these two expressions implies
\begin{equation}\label{det}
ad-bc=1\,.
\end{equation}

Substituting the ansatz (\ref{can}) into (\ref{eq1}), we obtain a system
of partial differential equations for the coefficient functions
$a$, $b$, $c$, and $d$:
\begin{eqnarray}
\label{eq:matching-coef}
&&a_u(2cu+dz)+b_u(cz+2dv)
-c_u(2au+bz)-d_u(az+2bv)=2f\,, \nonumber\\
&&a_z(2cu+dz)+b_z(cz+2dv)
-c_z(2au+bz)-d_z(az+2bv)+ad-bc=2g\,, \nonumber\\
&&a_z(2cu+dz)+b_z(cz+2dv)
-c_z(2au+bz)-d_z(az+2bv)+bc-ad=2h\,, \nonumber\\
&&a_v(2cu+dz)+b_v(cz+2dv)
-c_v(2au+bz)-d_v(az+2bv)=2k\,.
\end{eqnarray}
Here,
$a_u:=\partial a/\partial u$,
$a_v:=\partial a/\partial v$,
$a_z:=\partial a/\partial z$,
and similarly for the other coefficient functions.

The second and third equations are not independent. Indeed, subtracting
the third equation from the second and using (\ref{g-h}) and (\ref{det}),
we obtain an identity.

Using the explicit expressions for $f$, $g$, and $k$ in
(\ref{eq:fghk-closed}), the independent equations reduce to
\begin{eqnarray}
&&a_u(2cu+dz)+b_u(cz+2dv)
-c_u(2au+bz)-d_u(az+2bv)
=-4\left(\alpha^2z+2\alpha\beta v\right)F(\lambda),
\notag\\
&&a_z(2cu+dz)+b_z(cz+2dv)
-c_z(2au+bz)-d_z(az+2bv)
=4\left(\alpha^2u-\beta^2v\right)F(\lambda),
\notag\\
&&a_v(2cu+dz)+b_v(cz+2dv)
-c_v(2au+bz)-d_v(az+2bv)
=4\left(\beta^2z+2\alpha\beta u\right)F(\lambda),
\label{eq:reduced-abcd-system}
\end{eqnarray}
together with the algebraic condition (\ref{det}).

The structure of the right-hand sides of
(\ref{eq:reduced-abcd-system}) suggests the following ansatz:
\begin{eqnarray}
\label{eq:abcd-ansatz}
a(u,v,z)&=&A(\lambda)
-\left(\alpha^2u-\beta^2v\right)B(\lambda),
\nonumber\\
b(u,v,z)&=&-\left(\beta^2z+2\alpha\beta u\right)B(\lambda),
\nonumber\\
c(u,v,z)&=&-\left(\alpha^2z+2\alpha\beta v\right)B(\lambda),
\nonumber\\
d(u,v,z)&=&A(\lambda)
+\left(\alpha^2u-\beta^2v\right)B(\lambda).
\end{eqnarray}

Upon substituting this ansatz into
(\ref{eq:reduced-abcd-system}), all terms involving the derivatives
$A'(\lambda)$ and $B'(\lambda)$ cancel, and the three differential
equations reduce to a single algebraic condition,
\begin{equation}
\label{eq:A-rho-condition-2}
B\left(A-\frac{\lambda}{2}B\right)=2F(\lambda).
\end{equation}
The determinant condition (\ref{det}) gives, in addition,
\begin{equation}
\label{eq:A-rho-condition-1}
ad-bc=A^2-\frac{\lambda^2}{4}B^2=1.
\end{equation}

Thus, the ansatz (\ref{eq:abcd-ansatz}) reduces the system
(\ref{det}) and (\ref{eq:reduced-abcd-system}) to two algebraic
equations for the two unknown functions $A(\lambda)$ and $B(\lambda)$.
Their solution is
\begin{eqnarray}
A(\lambda)
&=&\frac{1-\lambda F(\lambda)}
{\sqrt{1-2\lambda F(\lambda)}}
=\frac{\lambda+\ln(1+\lambda)}
{2\sqrt{\lambda\,\ln(1+\lambda)}}\,,
\label{ALambda}\\
B(\lambda)
&=&\frac{2F(\lambda)}
{\sqrt{1-2\lambda F(\lambda)}}
=\frac{\lambda-\ln(1+\lambda)}
{\sqrt{\lambda^3\ln(1+\lambda)}}\,.
\notag
\end{eqnarray}

The resulting transformation is
\begin{eqnarray}\label{eq2}
y_\mu
&=&
\left[A(\lambda)-\left(\alpha^2u-\beta^2v\right)B(\lambda)\right]x_\mu
-\left(\beta^2z+2\alpha\beta u\right)B(\lambda)\,p_\mu,
\nonumber\\
\xi_\mu
&=&
-\left(\alpha^2z+2\alpha\beta v\right)B(\lambda)\,x_\mu
+\left[A(\lambda)+\left(\alpha^2u-\beta^2v\right)B(\lambda)\right]p_\mu.
\end{eqnarray}

This transformation maps the canonical action (\ref{ac2}) to the action
(\ref{ac1}), whose Poisson structure is the Snyder--de Sitter algebra
(\ref{SdS}).

\section{Inverse transformation}

Using the determinant condition (\ref{det}), we can immediately write the
inverse transformation as
\begin{eqnarray}\label{rel2}
x_\mu
&=&
\left[A(\lambda)+\left(\alpha^2u-\beta^2v\right)B(\lambda)\right]y_\mu
+\left(\beta^2z+2\alpha\beta u\right)B(\lambda)\,\xi_\mu,
\nonumber\\
p_\mu
&=&
\left(\alpha^2z+2\alpha\beta v\right)B(\lambda)\,y_\mu
+\left[A(\lambda)-\left(\alpha^2u-\beta^2v\right)B(\lambda)\right]\xi_\mu.
\end{eqnarray}

It remains to express the variables $u$, $v$, and $z$, defined below
(\ref{a1}), in terms of the canonical variables
\[
\bar u=\frac{y^2}{2},\qquad
\bar v=\frac{\xi^2}{2},\qquad
\bar z=y\cdot\xi.
\]
To this end, we first observe that (\ref{eq2}) implies
\begin{equation}\label{rel}
\alpha\,y_\mu+\beta\,\xi_\mu
=
\left(A(\lambda)-\frac{\lambda}{2}B(\lambda)\right)
\left(\alpha\,x_\mu+\beta\,p_\mu\right).
\end{equation}

Contracting both sides of (\ref{rel}) with $y^\mu$ and $\xi^\mu$,
respectively, and then using (\ref{eq2}) on the right-hand side, we obtain
\begin{eqnarray}\label{rel1}
\beta\,\bar z+2\alpha\,\bar u
&=&
\left(A(\lambda)-\frac{\lambda}{2}B(\lambda)\right)^2
\left(\beta\,z+2\alpha u\right),
\nonumber\\
\alpha\,\bar z+2\beta\,\bar v
&=&
\left(A(\lambda)-\frac{\lambda}{2}B(\lambda)\right)^2
\left(\alpha\,z+2\beta v\right).
\end{eqnarray}

It follows from (\ref{rel1}) that
\begin{equation}\label{lL}
\bar\lambda
=
\left(A(\lambda)-\frac{\lambda}{2}B(\lambda)\right)^2\lambda
=\ln(1+\lambda),
\end{equation}
where
\begin{equation}
\bar\lambda
=
2\alpha^2\bar u+2\beta^2\bar v+2\alpha\beta\bar z
=
\alpha^2y^2+\beta^2\xi^2+2\alpha\beta\,y\cdot\xi.
\end{equation}
Consequently,
\begin{equation}\label{rel3}
\lambda=e^{\bar\lambda}-1.
\end{equation}

It also follows from (\ref{rel1}) and (\ref{ALambda}) that
\begin{eqnarray}\label{rel4}
\alpha^2\bar u-\beta^2\bar v
&=&
\frac{\ln(1+\lambda)}{\lambda}
\left(\alpha^2u-\beta^2v\right),
\nonumber\\
\beta^2\bar z+2\alpha\beta\bar u
&=&
\frac{\ln(1+\lambda)}{\lambda}
\left(\beta^2z+2\alpha\beta u\right),
\nonumber\\
\alpha^2\bar z+2\alpha\beta\bar v
&=&
\frac{\ln(1+\lambda)}{\lambda}
\left(\alpha^2z+2\alpha\beta v\right).
\end{eqnarray}

Finally, substituting (\ref{rel3}) and (\ref{rel4}) into (\ref{rel2}), we
obtain
\begin{eqnarray}\label{rel5}
x_\mu
&=&
\left[C(\bar\lambda)
+\left(\alpha^2\bar u-\beta^2\bar v\right)D(\bar\lambda)\right]y_\mu
+\left(\beta^2\bar z+2\alpha\beta\bar u\right)D(\bar\lambda)\,\xi_\mu,
\nonumber\\
p_\mu
&=&
\left(\alpha^2\bar z+2\alpha\beta\bar v\right)D(\bar\lambda)\,y_\mu
+\left[C(\bar\lambda)
-\left(\alpha^2\bar u-\beta^2\bar v\right)D(\bar\lambda)\right]\xi_\mu,
\end{eqnarray}
where
\begin{eqnarray}
C(\bar\lambda)
&=&
\frac{e^{\bar\lambda}-1+\bar\lambda}
{2\sqrt{\left(e^{\bar\lambda}-1\right)\bar\lambda}}
=
1+\frac{\bar\lambda^2}{32}
+{\cal O}(\bar\lambda^3),
\nonumber\\
D(\bar\lambda)
&=&
\frac{e^{\bar\lambda}-1-\bar\lambda}
{\sqrt{\left(e^{\bar\lambda}-1\right)\bar\lambda^3}}
=
\frac12+\frac{\bar\lambda}{24}
+{\cal O}(\bar\lambda^2).\label{CD}
\end{eqnarray}

Equation (\ref{rel5}) provides an explicit all-order canonical representation of the
Snyder--de Sitter algebra that is regular in both parameters $\alpha$ and
$\beta$. This is the main technical result of the present work. The
representation is not unique and need not be the simplest possible one.
The freedom in constructing such representations is related to canonical
transformations, as discussed in our previous work
\cite{Kupriyanov:2025uen}.

\section{Poisson gauge algebra for Snyder--de Sitter space}

As an application of the canonical representation constructed above, we
consider the construction of a Poisson gauge theory on Snyder--de Sitter
space. Poisson gauge theory \cite{Kupriyanov:2021aet}, also known as Poisson
electrodynamics \cite{Kupriyanov:2023qot}, is a field-theoretical model
obtained as the semiclassical approximation of a noncommutative gauge
theory. In the semiclassical limit, the noncommutative gauge algebra
\begin{equation}
\left[\delta^{NC}_f,\delta^{NC}_g\right]
=\delta^{NC}_{-i[f,g]_\star}
\end{equation}
is replaced by the Poisson gauge algebra
\begin{equation}\label{pga}
\left[\delta_f,\delta_g\right]=\delta_{\{f,g\}}\,,
\end{equation}
where the Poisson bracket is obtained from the semiclassical limit of the
star commutator,
\begin{equation}
\{f,g\}
=\lim_{\hbar\to0}\frac{1}{i\hbar}[f,g]_\star\,.
\end{equation}
Poisson electrodynamics thus provides an effective field-theoretical
framework for investigating semiclassical aspects and solutions of
noncommutative gauge theories. For recent developments and physical
applications, see, e.g.,
\cite{Sharapov:2024bbu,Abla:2024wtr,Kurkov:2025soi,DiCosmo:2025mme,
Kurkov:2025abv,Sharapov:2026ife,Kurkov:2026qey}.

A consistent formulation of gauge theory on Snyder--de Sitter, or even on
Snyder space alone, remains an open problem. Here, we approach this problem
at the semiclassical level by constructing a Poisson gauge theory on
Snyder--de Sitter phase space. A crucial observation is that the Poisson
bracket of the coordinates in (\ref{SdS}) does not close on the coordinates
$x^\mu$ alone: it depends explicitly on both $x^\mu$ and their conjugate
momenta $p_\mu$. Consequently, the gauge field must be defined on the full
phase space. In other words, instead of a gauge potential $A_\mu(x)$
depending only on spacetime coordinates, we introduce a phase-space gauge
field
\begin{equation}
A_{\cal M}(X)=\big(A_\mu(x,p),A^\mu_\ast(x,p)\big)\,.
\end{equation}
This introduces additional degrees of freedom associated with the
momentum directions. An attempt to formulate a gauge theory on Snyder
space without these additional degrees of freedom leads to
nonassociativity \cite{Meljanac:2017grw}, whose physical interpretation
remains unclear.

Our goal is to construct gauge transformations
$\delta_f A_{\cal M}$ that realize the Poisson gauge algebra (\ref{pga})
for the Snyder--de Sitter Poisson structure (\ref{SdS}). To see the
origin of the difficulty, consider first a constant Poisson tensor
$\Theta=\Theta_0$. This situation is realized in the flat
$\alpha\to0$ and commutative $\beta\to0$ limits. In this case, the Jacobi
identity together with the standard Leibniz rule implies that the Poisson
gauge transformation can be written in the familiar form
\begin{equation}\label{gtcan}
\delta^{\mathrm{can}}_f A_{\cal M}
=\partial_{\cal M}f+\{A_{\cal M},f\}_0\,,
\end{equation}
where
\begin{equation}
\{A_{\cal M},f\}_0
=\Theta_0^{{\cal N}{\cal R}}
\partial_{\cal N}A_{\cal M}\,
\partial_{\cal R}f\,.
\end{equation}

For a coordinate-dependent Poisson tensor $\Theta(X)$, however, the
standard Leibniz rule for the Poisson bracket and ordinary derivatives is
no longer sufficient. In particular,
\begin{equation}
\partial_{\cal M}\{f,g\}
\neq
\{\partial_{\cal M}f,g\}
+\{f,\partial_{\cal M}g\}\,.
\end{equation}
Therefore, additional terms are required in (\ref{gtcan}) to compensate
for this deviation. In general, constructing these correction terms is
nontrivial and can be approached using homotopy algebras
\cite{Kup27} or symplectic embeddings \cite{Kupriyanov:2021cws}.

The Snyder--de Sitter case is special because its Poisson structure is
symplectic. Consequently, Darboux's theorem guarantees the existence of
local coordinates in which the Poisson tensor is constant. In the present
work, we have constructed such a transformation explicitly in
(\ref{rel5}). This allows us to formulate the gauge transformations first
in the canonical Darboux coordinates and then transform them back to the
original phase-space variables.

Indeed, in the Darboux coordinates $Y^{\cal M}$, we define
\begin{equation}
\delta_f A_{\cal M}(Y)
=
\partial^Y_{\cal M}f(Y)
+\{A_{\cal M}(Y),f(Y)\}_0\,,
\qquad
\partial^Y_{\cal M}
=\frac{\partial}{\partial Y^{\cal M}}\,.
\end{equation}
Using the canonical representation (\ref{rel5}), we can then return to
the original Snyder--de Sitter variables $X^{\cal M}$ and obtain
\begin{equation}\label{gtSdS}
\delta_f A_{\cal M}(X)
=
\Gamma^{{\cal N}}_{\cal M}(X)\,
\partial_{\cal N}f
+\{A_{\cal M}(X),f(X)\}\,,
\end{equation}
where
\begin{equation}\label{Gamma}
\Gamma^{{\cal N}}_{\cal M}(X)
=
\left.
\frac{\partial X^{\cal N}}
{\partial Y^{\cal M}}
\right|_{Y=Y(X)}\,.
\end{equation}

The explicit form of the matrix $\Gamma(X)$ is given in the Appendix. By
construction, the transformations (\ref{gtSdS}) close according to the
Poisson gauge algebra (\ref{pga}) with the Snyder--de Sitter Poisson
structure (\ref{SdS}). The derivation of the corresponding Poisson gauge
transformations in the framework of symplectic embeddings was presented
in \cite{Kupriyanov:2025uen}. We emphasize, however, that the matrix
$\Gamma(X)$ obtained in \cite{Kupriyanov:2025uen} was irregular in the
parameter $\beta$. In contrast, the matrix (\ref{Gamma}) constructed here
is regular in both $\alpha$ and $\beta$. This regularity follows directly
from the regularity of the canonical representation (\ref{rel5}).

The corresponding Poisson field strength is defined by
\begin{equation}
{\cal F}_{{\cal M}{\cal N}}
=
\Gamma^{{\cal K}}_{\cal M}\,
\partial_{\cal K}A_{\cal N}
-\Gamma^{{\cal K}}_{\cal N}\,
\partial_{\cal K}A_{\cal M}
+\{A_{\cal M},A_{\cal N}\}\,.
\end{equation}
It transforms covariantly under (\ref{gtSdS}),
\begin{equation}
\delta_f{\cal F}_{{\cal M}{\cal N}}
=
\{f,{\cal F}_{{\cal M}{\cal N}}\}\,.
\end{equation}
One can therefore proceed to construct gauge-covariant Lagrangians and
gauge-invariant actions. The physical interpretation and implications of
such a gauge theory, however, will be investigated in future work.

We note that the main conceptual issue concerns the physical
interpretation of the additional phase-space degrees of freedom. Even in
the simultaneous flat and commutative limit
$(\alpha,\beta)\to(0,0)$, the resulting theory is formulated on phase
space rather than on spacetime alone. In the present construction this
corresponds to a theory in $8$ dimensions, with gauge transformations
given by (\ref{gtcan}). Understanding how this phase-space formulation
is related to standard Maxwell electrodynamics in $4$ dimensions is an
important question that deserves further investigation.

Further applications of the canonical representation constructed in this
work may be explored in different aspects of classical and quantum
mechanics; see, for example,
\cite{Lukierski:2022yni,Pachol:2024hiz}.

\section*{Acknowledgments}
We appreciate the support of the Funda\c{c}c\~ao de Amparo a Pesquisa do Estado de São Paulo
(FAPESP, Brazil), Grant numbers: 2024/04134 - 6 and 2024/23831m- 0. KVG also acknowledges financial support from the Concelho Nacional de Pesquisa (CNPq, Brazil), grant number: 305132/2024 - 5. This study was financed in part by the Coordenação de Aperfeiçoamento de Pessoal de Nível Superior - Brasil (CAPES) - Finance Code 001.

\section*{Appendix}

The matrix $\Gamma(X)$ defined in (\ref{Gamma}), can be represented as
\begin{equation}    \label{def}
  \Gamma^{\cal N}_{\cal M}(X) = \left.
\frac{\partial X^{\cal N}}
{\partial Y^{\cal M}}
\right|_{Y=Y(X)}=\begin{pmatrix} \displaystyle
\Gamma^\nu_\mu(x,p) & \Gamma^{D+\nu}_\mu(x,p) \\
\displaystyle \Gamma^\mu_{D+\nu}(x,p) & \displaystyle
\Gamma^{D+\mu}_{D+\nu}(x,p)
\end{pmatrix}.
\end{equation}
Recall that the functions $a(u,v,z)$, $b(u,v,z)$, $c(u,v,z)$, and
$d(u,v,z)$ are defined in (\ref{eq:abcd-ansatz}), while
$C(\bar\lambda)$ and $D(\bar\lambda)$ are given in (\ref{CD}). The first
block of the matrix $\Gamma(X)$ is therefore
\begin{align}
\Gamma^\nu_\mu(x,p)&={} d\,\delta^\nu{}_\mu
\nonumber\\
&+ \alpha^2 \Bigg[D\left(\ln(1+\lambda)\right) + 2C'\left(\ln(1+\lambda)\right) + 2\frac{\ln(1+\lambda)}{\lambda}
\left( \alpha^2u-\beta^2v \right) D'\left(\ln(1+\lambda)\right)
\Bigg] \nonumber\\
&\qquad\times
\left(a x^\nu+b p^\nu \right) \left( a x_\mu+b p_\mu \right)
\nonumber\\
&+2\alpha\beta \Bigg[C'\left(\ln(1+\lambda)\right) + \frac{\ln(1+\lambda)}{\lambda} \left(\alpha^2u-\beta^2v
\right) D'\left(\ln(1+\lambda)\right) \Bigg] \nonumber\\
&\qquad\times
\left(a x^\nu+b p^\nu \right) \left( c x_\mu+d p_\mu \right) \nonumber\\
&+ \Bigg[2\alpha\beta D\left(\ln(1+\lambda)\right)
+ 2\alpha^2 \frac{\ln(1+\lambda)}{\lambda} \left( \beta^2z+2\alpha\beta u\right)D'\left(\ln(1+\lambda)\right)
\Bigg] \nonumber\\
&\qquad\times \left(c x^\nu+d p^\nu\right)\left(a x_\mu+b p_\mu
\right) \nonumber\\
&+ \Bigg[\beta^2 D\left(\ln(1+\lambda)\right) + 2\alpha\beta
\frac{\ln(1+\lambda)}{\lambda} \left(\beta^2z+2\alpha\beta u
\right) D'\left(\ln(1+\lambda)\right)
\Bigg] \nonumber\\
&\qquad\times \left(c x^\nu+d p^\nu \right)\left( c x_\mu+d p_\mu\right).
\label{eq:Gamma-xy-X}
\end{align}

The second block is given by
\begin{align}
\Gamma^{D+\nu}_\mu(x,p) &={} -b\,\eta^{\nu\mu}
+ 2\alpha\beta \Bigg[ C'\left(\ln(1+\lambda)\right) + \frac{\ln(1+\lambda)}{\lambda}\left(\alpha^2u-\beta^2v \right)D'\left(\ln(1+\lambda)\right) \Bigg]\nonumber\\
&\qquad\times
\left(a x^\nu+b p^\nu \right)\left(a x^\mu+b p^\mu \right) \nonumber\\
&+ \beta^2 \Bigg[ 2C'\left(\ln(1+\lambda)\right) + 2\frac{\ln(1+\lambda)}{\lambda} \left( \alpha^2u-\beta^2v \right)
D'\left(\ln(1+\lambda)\right) - D\left(\ln(1+\lambda)\right)
\Bigg] \nonumber\\
&\qquad\times \left( a x^\nu+b p^\nu \right) \left(c x^\mu+d p^\mu
\right)\nonumber\\
&+\Bigg[
\beta^2 D\left(\ln(1+\lambda)\right) + 2\alpha\beta
\frac{\ln(1+\lambda)}{\lambda} \left( \beta^2z+2\alpha\beta u \right) D'\left(\ln(1+\lambda)\right)
\Bigg] \nonumber\\
&\qquad\times \left( c x^\nu+d p^\nu \right)\left( a x^\mu+b p^\mu
\right) \nonumber\\
&+ 2\beta^2\frac{\ln(1+\lambda)}{\lambda}
\left( \beta^2z+2\alpha\beta u \right) D'\left(\ln(1+\lambda)\right) \nonumber\\
&\qquad\times
\left( c x^\nu+d p^\nu \right) \left( c x^\mu+d p^\mu \right).
\label{eq:Gamma-xxi-X}
\end{align}

The third block is
\begin{align}
\Gamma^\mu_{D+\nu}(x,p)  &={}
-c\,\eta_{\nu\mu} \nonumber\\
&+ 2\alpha^2 \frac{\ln(1+\lambda)}{\lambda}
\left(\alpha^2z+2\alpha\beta v \right)
D'\left(\ln(1+\lambda)\right) \nonumber\\
&\qquad\times
\left(a x_\nu+b p_\nu \right)\left( a x_\mu+b p_\mu \right)
\nonumber\\
&+ \Bigg[\alpha^2 D\left(\ln(1+\lambda)\right) + 2\alpha\beta
\frac{\ln(1+\lambda)}{\lambda} \left( \alpha^2z+2\alpha\beta v \right) D'\left(\ln(1+\lambda)\right) \Bigg]\nonumber\\
&\qquad\times \left(a x_\nu+b p_\nu \right) \left(c x_\mu+d p_\mu \right) \nonumber\\
&+ \alpha^2 \Bigg[2C'\left(\ln(1+\lambda)\right)-
2\frac{\ln(1+\lambda)}{\lambda}\left(\alpha^2u-\beta^2v\right)
D'\left(\ln(1+\lambda)\right)- D\left(\ln(1+\lambda)\right)
\Bigg] \nonumber\\
&\qquad\times \left(c x_\nu+d p_\nu\right)\left(a x_\mu+b p_\mu \right)\nonumber\\
&+2\alpha\beta \Bigg[C'\left(\ln(1+\lambda)\right) -
\frac{\ln(1+\lambda)}{\lambda} \left(
\alpha^2u-\beta^2v\right) D'\left(\ln(1+\lambda)\right)
\Bigg] \nonumber\\
&\qquad\times \left( c x_\nu+d p_\nu
\right) \left( c x_\mu+d p_\mu \right).
\label{eq:Gamma-py-X}
\end{align}

Finally, the fourth block is
\begin{align}
\Gamma^{D+\mu}_{D+\nu}(x,p)&={} a\,\delta_\nu{}^\mu \nonumber\\
&+ \Bigg[ \alpha^2 D\left(\ln(1+\lambda)\right) + 2\alpha\beta \frac{\ln(1+\lambda)}{\lambda} \left(
\alpha^2z+2\alpha\beta v \right) D'\left(\ln(1+\lambda)\right)
\Bigg] \nonumber\\
&\qquad\times \left(a x_\nu+b p_\nu \right)\left(a x^\mu+b p^\mu \right)\nonumber\\
&+\Bigg[ 2\alpha\beta D\left(\ln(1+\lambda)\right) + 2\beta^2 \frac{\ln(1+\lambda)}{\lambda}\left(
\alpha^2z+2\alpha\beta v \right) D'\left(\ln(1+\lambda)\right)
\Bigg] \nonumber\\
&\qquad\times \left(a x_\nu+b p_\nu \right) \left( c x^\mu+d p^\mu \right)
\nonumber\\
&+ 2\alpha\beta \Bigg[ C'\left(\ln(1+\lambda)\right) - \frac{\ln(1+\lambda)}{\lambda} \left(
\alpha^2u-\beta^2v \right) D'\left(\ln(1+\lambda)\right) \Bigg]
\nonumber\\
&\qquad\times \left( c x_\nu+d p_\nu \right) \left( a x^\mu+b p^\mu \right)
\nonumber\\
&+ \beta^2 \Bigg[ D\left(\ln(1+\lambda)\right) + 2C'\left(\ln(1+\lambda)\right)  - 2\frac{\ln(1+\lambda)}{\lambda} \left(\alpha^2u-\beta^2v
\right) D'\left(\ln(1+\lambda)\right) \Bigg] \nonumber\\
&\qquad\times \left( c x_\nu+d p_\nu \right) \left( c x^\mu+d p^\mu \right).
\label{eq:Gamma-pxi-X}
\end{align}
All expressions containing $\ln(1+\lambda)/\lambda$ are understood by
continuity at $\lambda=0$, where
$\ln(1+\lambda)/\lambda=1+\mathcal{O}(\lambda)$.
Consequently, the matrix $\Gamma(X)$ is regular at
$(\alpha,\beta)=(0,0)$. Equations~\eqref{eq:Gamma-xy-X}--\eqref{eq:Gamma-pxi-X}
constitute the matrix $\Gamma^{\cal N}_{\cal M}(X)$ entirely in terms of the
original Snyder--de Sitter phase-space variables $x^\mu$ and $p_\mu$.

\end{document}